\documentclass[twoside]{article}
\usepackage{fleqn,espcrc2}
\usepackage{epsfig}

\title{Atmospheric, Solar, and CHOOZ neutrinos: 
             a global three generation analysis}

\author{
	G.L.\ Fogli$^{\rm a}$,\
	E.\ Lisi$^{\rm a}$,
	A.\ Marrone$^{\rm a}$,
        D.\ Montanino$^{\rm b}$, and
	A.\ Palazzo
\address{
	Dipartimento di Fisica and Sezione INFN di Bari, 
	Via Amendola 173, 70126 Bari, Italy\\
	$^{\rm b}$ Dipartimento di Scienza dei Materiali
	dell'Universit\`a di Lecce, Via Arnesano, 73100 Lecce, Italy}
}

\begin{document}

\begin{abstract}
We perform a global three generation analysis of the current solar and
atmospheric evidence in favor of neutrino oscillations. We also include the
negative results coming from CHOOZ to constrain the $\nu_e$ mixing. We study
the zones of mass--mixing oscillations parameters compatible with all the data.
It is shown that almost pure $\nu_\mu\leftrightarrow \nu_\tau$ oscillations are
required to explain the atmospheric neutrino anomaly and almost pure
$\nu_1\leftrightarrow \nu_2$ oscillations to account for the solar neutrino
deficit. 
\end{abstract} \maketitle

\section{Introduction}

The evidence for non zero neutrino mass and mixing is, at present, the only
solid experimental hint for physics beyond the Standard Model. Flavor
oscillations~\cite{Pont} are the privileged tool to explore the neutrino mass
and mixing parameter space. At present, there are three experimental
indications in favor of neutrino flavor oscillations: 1) the evidence for
$\nu_e$ appearance from a $\nu_\mu$ beam in the Liquid Scintillator Neutrino
Detector (LSND)~\cite{LSND}; 2) the evidence for $\nu_e$ disappearance in the
solar neutrino flux~\cite{NuAs}; 3) the strong evidence for $\nu_\mu$
suppression in the atmospheric neutrino flux, together with the evidence of
$L$--dependence of such suppression, in SuperKamiokande (SK)~\cite{SK00}, as
well as in MACRO and Soudan 2~\cite{MaSo}.

Any of the above pieces of evidence requires a different mass scale: $\Delta
m^2 \sim O(1\, {\rm eV}^2)$ for LSND, $\Delta m^2 \sim O(10^{-3}\, {\rm eV}^2)$
for atmospheric neutrinos, and $\Delta m^2 \leq 10^{-4}\, {\rm eV}^2$ for solar
neutrinos. To account for all of them we need at least one sterile neutrino.
Anyway, the evidence coming from LSND is, at the moment, controversial. For
this reason, waiting for an independent confirmation of the LSND result, we
prefer to discard this datum and to analyze only the solar and atmospheric
evidence of oscillation in a ``standard'' scenario with three active neutrinos.
In addition, we consider also the negative evidence coming from CHOOZ
\cite{CH00}. Such 1 km--baseline reactor experiment has not found any evidence
for $\nu_e$ disappearance. As we will see, this negative result has a strong
impact in constraining the $3\nu$ parameter space. (Similar conclusions have
also been derived by Gonzales--Garcia {\em et al.}~\cite{Go00})

\section{The standard $3\nu$ framework}

Flavor eigenstates are related to mass eigenstates through the mixing matrix
$U$:
\begin{equation}
\nu_\alpha = \sum_{i=1}^3 U_{\alpha i} \nu_i, ,
\label{eq:flavmass}
\end{equation}
where $\alpha = e, \mu, \tau$. We stick in the ``one mass scale dominance
hypothesis'', i.e., $\delta m^2 \equiv m_2^2 - m_1^2 \ll m^2 \equiv |m_3^2 -
m_{1,2}^2|$. In this hypothesis -- that will be proved {\em a posteriori} -- CP
violating effects are unobservable and the matrix elements $U_{\alpha i}$ can
be considered, without loss of generality, real. Moreover, atmospheric (and,
eventually, CHOOZ) neutrino oscillations probe the flavor composition of the
``lone'' state $\nu_3$ and can be described by the subspace $(m^2, U_{e3},
U_{\mu 3}, U_{\tau 3})$. Conversely, solar neutrino oscillations can probe the
mass composition of the $\nu_e$ and can be described by the subspace $(\delta
m^2, U_{e1}, U_{e2}, U_{e3})$. The only common parameter that can be probed
both by solar and atmospheric neutrino oscillations is $U_{e 3}$. 

\begin{figure*}[t!]
\begin{center}
\epsfig{bbllx = 150, bblly = 180, bburx = 500, bbury = 720, 
        height = 10.6truecm, figure = 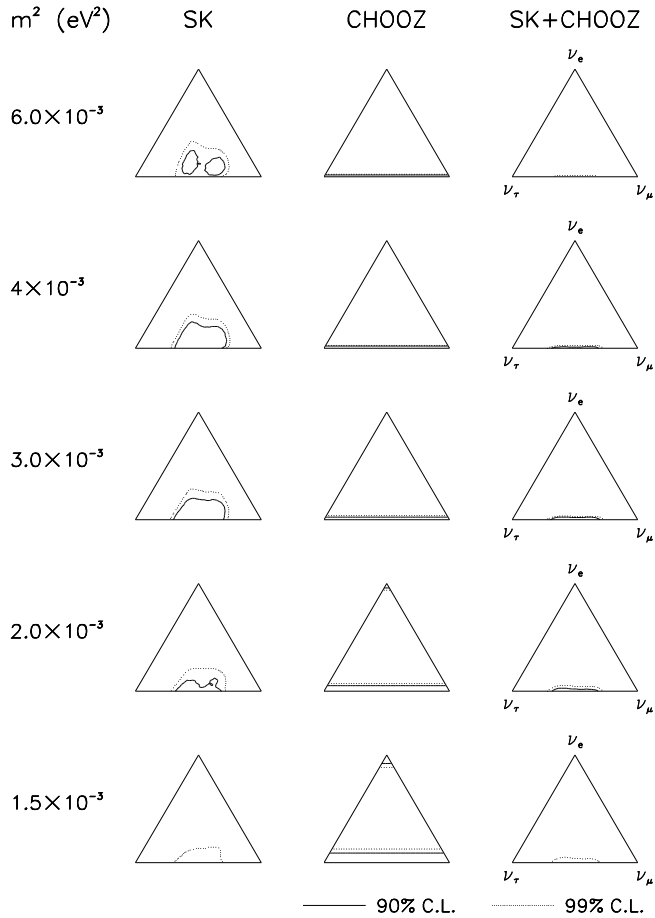}%
\end{center}
\caption{\label{fig:atmos} 
Allowed zones at 90\% (99\%) C.L. of SK atmospheric data (first column), CHOOZ
(second column) and combined.}%
\end{figure*}

\section{3$\nu$ Atmospheric and CHOOZ analysis}

We have performed an updated analysis~\cite{Ma01} of the latest (70.5 kTy) data
from SK~\cite{SK00} and CHOOZ~\cite{CH00}. The details of the analysis can be
found in~\cite{Fo99}. The SK data include 55 zenith bins: 10+10 bins for the
subGeV $e$+$\mu$ events, 10+10 bins for the multiGeV $e$+$\mu$ events, and 5+10
bins for the upward stopping (US) and through-going $\mu$ events. For CHOOZ, we
use the 14 experimental bins.

\begin{figure*}[t!]
\begin{center}
\begin{tabular}{c@{\hspace*{0.8truecm}}c}
\mbox{\epsfig{bbllx = 90 , bblly = 180, bburx = 550, bbury = 750,
              height = 9.2truecm, figure = 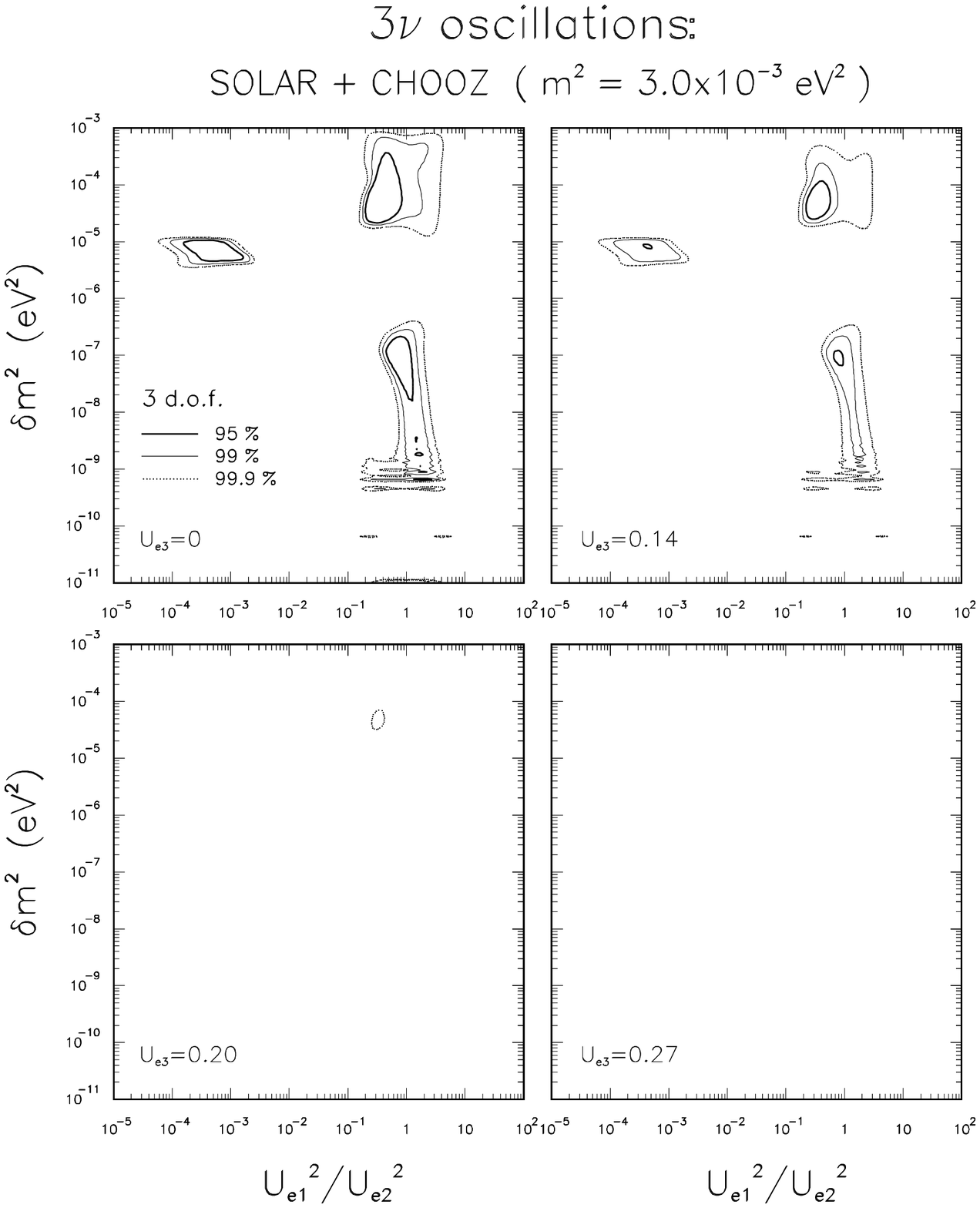}}&
\mbox{\epsfig{bbllx = 90 , bblly = 180, bburx = 550, bbury = 750,
              height = 9.2truecm, figure = 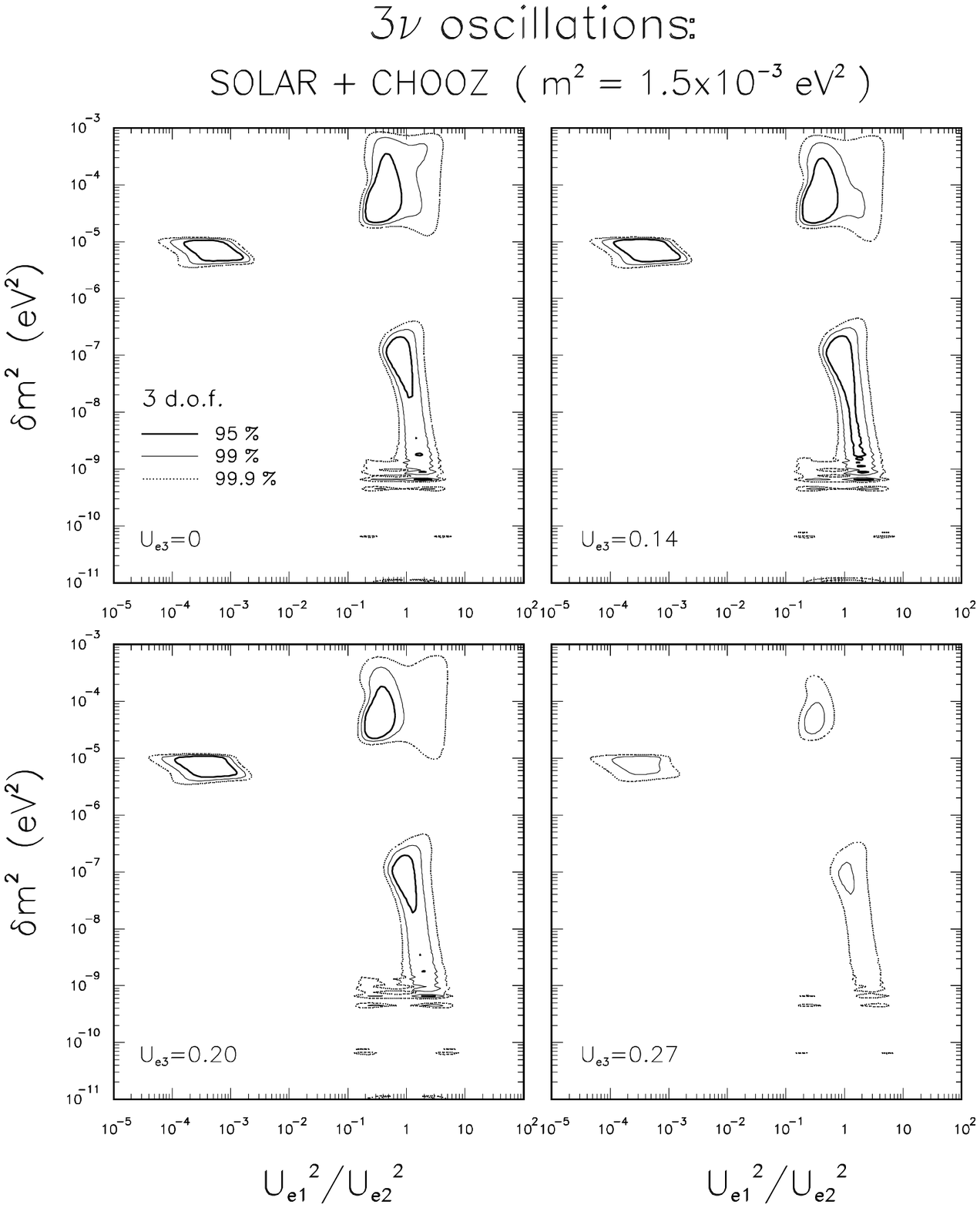}}
\end{tabular}
\end{center}
\caption{\label{fig:solar} 
Allowed zones at 95\%, 99\% and 99.9\% C.L. for the combination of all solar
$\nu$ data and CHOOZ, for four representative values of $U_{e3}$ and two
representative values of $m^2$.}%
\end{figure*}

The result of the analysis is shown in Fig. \ref{fig:atmos}, where the
unitarity triangle introduced in~\cite{Fo96} has been used. A point inside each
triangle in Fig. \ref{fig:atmos} represents a generic combination of the flavor
eigenstates (for a fixed $m^2$), the mixing matrix elements $U_{e 3}^2$,
$U_{\mu 3}^2$, and $U_{\tau 3}^2$ being identified with the projected heights
onto the $\nu_\tau$--$\nu_\mu$, $\nu_\tau$--$\nu_e$, and $\nu_\mu$--$\nu_e$
``sides'' respectively. Using a well known property of the triangles with equal
sides, the unitarity relation $U_{e 3}^2 + U_{\mu 3}^2 + U_{\tau 3}^2 = 1$ is
thus satisfied.

The triangles on the left of Fig. \ref{fig:atmos} show the zones allowed using
SK atmospheric data only, for five representative values of $m^2$. The absence
of $\nu_e$ distorsion in the atmospheric flux tends excludes pure $\nu_\mu
\leftrightarrow \nu_e$ oscillations, although a moderate mixing of $\nu_e$ is
still allowed by SK data. The best fit is reached for $m^2 \simeq 3 \times
10^{-3}$ eV$^2$ and pure maximal $\nu_\mu \leftrightarrow \nu_\tau$ mixing
($U_{\mu 3}^2 \simeq U_{\tau 3}^2 \simeq 1/2$ and $U_{e 3}^2 \simeq 0$).

The middle column of triangles show the zones allowed by CHOOZ. In this
experiment no signal of $\nu_e$ disappearance has been found. This is crucial
in constraining the mixing between $\nu_3$ and $\nu_e$. The zones allowed by
CHOOZ are shaped as horizontal strips near the bottom (corresponding to small
values of $U_{e 3}^2$) and near the $\nu_e$ corner (corresponding to $U_{e 3}^2
\simeq 1$). The last solution is, however, incompatible both with solar and
atmospheric neutrino oscillations.

The triangles on the right show the combined SK+CHOOZ analysis. Only a very
small zone, around pure maximal $\nu_\mu \leftrightarrow \nu_\tau$
oscillations, is allowed. Solutions disappear at 99\% C.L. for $m^2 \geq 6
\times 10^{-3}$ eV$^2$ and $m^2 \leq 1.5 \times 10^{-3}$ eV$^2$. The best fit
is reached again for $m^2 \simeq 3 \times 10^{-3}$ eV$^2$ and pure maximal
$\nu_\mu \leftrightarrow \nu_\tau$ mixing. In particular, the upper limit on
$U_{e 3}$ is $\simeq 0.3$ at 99\% C.L.

\section{3$\nu$ Solar and CHOOZ analysis}

In this Section we present an updated analysis~\cite{Pa01} of the solar
neutrino data (total rates and SK Day and Night recoil spectrum, as presented
at the {\em Neutrino 2000} conference~\cite{Nu2000}), as well  as of the latest
theoretical solar $\nu$ fluxes and uncertainties~\cite{BaPi}. The details of
the analysis can be found in~\cite{Pa99}. Moreover, we have included the CHOOZ
constraints on $U_{e 3}$. In this case, it is necessary to fix the value of
$m^2$, since the leading oscillations in CHOOZ are driven by the higher mass
gap ($m^2$).

In Figure 2 we show the results of the analysis for $m^2 = 3 \times 10^{-3}$
and $1.5 \times 10^{-3}$ eV$^2$ (corresponding, respectively, to the best fit
and to the lowest value allowed by atmospheric $\nu$ oscillations), and for
increasing values of $U_{e 3}$. The case $U_{e 3} = 0$ corresponds to the usual
$2\nu$ analysis, with the identification $U_{e 1}^2 / U_{e 2}^2 \equiv
\tan^2\theta$. The best fit is reached for $U_{e 3} = 0$, $U_{e 1}^2 / U_{e
2}^2 \simeq 0.36$, and $\delta m^2 \simeq 4.7 \times 10^{-5}$ eV$^2$. The
solution at small mixing angle is disfavored by the lack of the evidence of
distorsion in the SK spectrum. Solar neutrinos alone prefer $U_{e 3} = 0$,
although the upper limit on $U_{e 3}$ is weak ($U_{e 3} \leq 0.8$
\cite{Pa99}). 

For increasing values of $U_{e 3}$ the solutions rapidly disappear because they
become incompatible with CHOOZ. In particular, the upper limit at 99\% C.L. on
$U_{e 3}$ is $\simeq 0.3$ for $m^2 = 1.5 \times 10^{-3}$ eV$^2$. The inclusion
of CHOOZ in the analysis also strongly constrain the upper value of $\delta
m^2$: $\delta m^2 \leq 7 \times 10^{-4}$ eV$^2$ at 99\% C.L. For such high
values of $\delta m^2$ the one mass scale dominance is valid only
approximatively. For this reasons the subleading effects of nonzero $\delta
m^2$ in CHOOZ and finite $m^2$ in solar analysis have been taken into account
in the analysis.

\section{Conclusions}

We have presented an updated $3\nu$ analysis of the atmospheric neutrino
anomaly and the solar neutrino deficit together with the negative evidence
coming from CHOOZ. Atmospheric neutrinos prefer almost pure $\nu_\mu
\leftrightarrow \nu_\tau$ oscillations with large mixing between $\nu_\mu$ and
$\nu_\tau$ and $m^2 \in [1.5,6] \times 10^{-3}$ eV$^2$. In particular, the best
fit is reached for maximal pure $\nu_\mu$--$\nu_\tau$ mixing. Solar neutrinos
still allow a multiplicity of solutions (with $\delta m^2 \leq 7 \times
10^{-4}$ eV$^2$), although large $\nu_1$--$\nu_2$ mixing is preferred. The
combined analysis with CHOOZ strongly constrains the $U_{e 3}$ mixing ($U_{e
3}\leq 0.3$). In particular, a theoretical attractive scenario, called
``bimaximal mixing'' ($U_{e 1}^2 = U_{e 2}^2 = 1/2 = U_{\mu 3}^2 = U_{\tau
3}^2$, $U_{e 3}^2 = 0$)~\cite{Ba98} is allowed. The goal for the next
generations of experiments is to constrain more tightly the parameter space(s)
and eventually to check (or disprove) non--standard solutions of the current
evidences of neutrino oscillations.

\section*{Aknowledgments}

D.M. thanks the organizers of the ``Rencontres de Moriond'' for kind
hospitality. This work is co--financed by the INFN and by Italian Ministero
dell'Universit{\`a} e della Ricerca Scientifica e Tecnologica (MURST) within
the ``Astroparticle Physics'' project.



\end{document}